\documentclass[12pt]{article}
\usepackage{amsfonts,amsmath}

\def\be{\begin{equation}}
\def\ee{\end{equation}}
\def\bea{\begin{eqnarray}}
\def\eea{\end{eqnarray}}

\newcommand{\sect}[1]{\setcounter{equation}{0}\section{#1}}
\newcommand{\subsect}[1]{\subsection{#1}}

\newcommand{\bq}{\mathbf{q}}
\newcommand{\bp}{\mathbf{p}}
\newcommand{\cM}{{\mathcal M}}

\newcommand\Om\Omega

\newcommand{\De}{\Delta}
\newcommand\dif[2]{\frac{\pd #1}{\pd #2}}
\newcommand{\pd}{\partial}
\newcommand\minus\backslash
\newcommand{\Kep}{_{\mathrm{KC}}}
\newcommand{\Harm}{_{\mathrm{ho}}}
\newcommand{\cD}{{\mathcal D}}
\newcommand{\e}{{\mathrm e}}

\def\RR{\mathbb{R}}
\def\cte{\alpha}
\def\bU{\mathcal{B}}

\def\rr{{\hat r}}
\def\rrr{r}
\def\ra{{\tilde r}}

\def\1{\'{\i}}

\def\k{{\kappa}}

\def\kk{k}

\def\dd{{\rm d}}
\def\>#1{{\mathbf#1}}

\begin{document}

 \
 \bigskip

\noindent
{\Large{\bf{Bertrand spacetimes as Kepler/oscillator  potentials}}}

\bigskip

\begin{center}
{\'Angel Ballesteros~$^a$, Alberto Enciso~$^b$, Francisco J. Herranz~$^c$
and Orlando Ragnisco~$^d$}
\end{center}

\noindent
{$^a$ Depto.~de F\'\i sica, Facultad de Ciencias, Universidad de Burgos,
09001 Burgos, Spain\\ ~~E-mail: angelb@ubu.es\\[10pt]
$^b$ Depto.~de F{\'\i}sica Te\'orica II,   Universidad Complutense,   28040 Madrid,
Spain\\ ~~E-mail: aenciso@fis.ucm.es\\[10pt]
$^c$ Depto.~de F\'\i sica,  Escuela Polit\'ecnica Superior, Universidad de Burgos,
09001 Burgos, Spain \\ ~~E-mail: fjherranz@ubu.es\\[10pt]
$^d$ Dipartimento di Fisica,   Universit\`a di Roma Tre and Instituto Nazionale di
Fisica Nucleare sezione di Roma Tre,  Via Vasca Navale 84,  00146 Roma, Italy  \\
~~E-mail: ragnisco@fis.uniroma3.it}

\medskip
\medskip

\begin{abstract}
\noindent
Perlick's classification of  $(3+1)$-dimensional spherically symmetric and static  spacetimes $(\mathcal M,\eta=-{\frac{1}{V}}\,\dd t^2+g)$ for which the classical Bertrand theorem holds  [Perlick V  1992   {\it Class.\ Quantum Grav.}  {\bf 9}    1009]  is revisited.   For any  Bertrand spacetime  $(\mathcal M,\eta)$ the term $V(r)$ is proven to be either the intrinsic Kepler--Coulomb or the harmonic oscillator potential on its associated Riemannian 3-manifold $(M,g)$. Among the latter 3-spaces $(M,g)$ we explicitly identify the three classical Riemannian spaces of constant curvature, a generalization of a Darboux space and the Iwai--Katayama spaces generalizing the MIC--Kepler and  Taub-NUT problems. The key dynamical role played by the Kepler and oscillator potentials in Euclidean space is thus extended to a wide class of 3-dimensional curved spaces.
 \end{abstract}

\bigskip\bigskip\bigskip\bigskip



\newpage


\sect{Introduction}

It is a well known, but nonetheless quite remarkable fact that all the bounded trajectories in  Euclidean
space associated with the Kepler--Coulomb (KC) potential  $V(r)=\alpha/r+\beta$ are periodic, so that they yield close orbits, i.e.\
ellipses. This is in striking  contrast e.g.\ with the gravitational 3-body problem,
which is known to possess extremely complicated bounded orbits. Similarly,
the orbits of the harmonic oscillator potential $V(r)=\omega^2\,r^2+\delta$ are all
closed. The content of the celebrated theorem of Bertrand~\cite{Be1873}
is that, under certain technical conditions, the KC and harmonic oscillator potentials
are the only spherically symmetric ones  for which
all the bounded trajectories of the Hamiltonian
$H(\bp,\bq)=\bp^2+V(|\bq|)$ are periodic.
The proof of Bertrand's theorem is not difficult, and has actually been
included (with various levels of rigor) in several textbooks. Nevertheless, this classical result provides a  wealth of subtle connections with
different areas of Physics, and still attracts
considerable attention, mainly motivated by its deep connection with accidental  degeneracy
and with superintegrability in quantum and classical mechanics~\cite{Za02,Fe04,Gr61,MR92}.

Some years ago,  in a remarkable paper, Perlick~\cite{perlick} showed that Bertrand's theorem also
arises naturally in General Relativity. Indeed, let us consider a spherically symmetric static spacetime
$(\cM,\eta)$ which is the Lorentzian warped product of the line by a
Riemannian 3-dimensional (3D) manifold $(M,g)$:
\be
\eta=-{\frac{1}{V}}\,\dd t^2+g\,.
\label{bs}
\ee
Of course, the  warping function ${V}\in C^\infty(M)$ must be strictly positive. It can be
easily seen that the projection of each timelike geodesic on a constant
time leaf $M_0=M\times\{t_0\}$ is in fact a trajectory of the natural
Hamiltonian $H(\bp,\bq)=\|\bp\|_{T^*M_0}^2+V(\bq)$. This relationship between an
autonomous natural Hamiltonian flow on a $3$D manifold with lower bounded
potential and the timelike geodesics of a Lorentzian causal manifold can
be understood as the Lorentzian analog of the introduction of the Jacobi
metric in classical mechanics \cite{AM}. Consequently, by a {\em
trajectory} in $(\cM,\eta)$ we shall mean the projection of a timelike
geodesic onto $M_0$.

Following Perlick, we shall say that a domain $\bU$ of a smooth $(3+1)$D Lorentzian
spacetime $(\cM,\eta)$ (possibly the whole space) is {\em a Bertrand spacetime}
provided that:
\begin{enumerate}
\item $(\cM,\eta)$ is spherically symmetric and static, the domain $\bU$ (possibly minus one or two points)  is diffeomorphic to a
product manifold $]r_1,r_2[\times \mathbb S^2\times \RR$ and the metric
$\eta$ in $\bU$ takes the form
\be
\eta=-{\frac{1}{V(r)}}\,\dd t^2+g(r)^2\dd r^2+r^2\big(\dd\vartheta^2+\sin^2\vartheta\,\dd\varphi^2\big).
\label{aaa}
\ee

\item There is a circular ($r=\text{constant}$) trajectory passing through every point of $\bU$.

\item Circular trajectories are stable under small perturbations of the initial conditions.
\end{enumerate}
Perlick \cite{perlick} posed and solved the problem of classifying all
Bertrand spacetimes, finding that the only possibilities are {\it three} multiparametric families of spacetimes, called hereafter of type I and II$_\pm$, which are  explicitly   given by
\bea
{\mbox {Type  I} }: && \dd
s^2=  -\frac{\dd t^2}{G+ \sqrt{ \rrr^{-2} +K} }+\frac{ \dd\rrr^2 }{\beta^2\left( 1+K\rrr^2\right) }
+\rrr^2(\dd\vartheta^2+\sin^2\vartheta\,\dd\varphi^2)
\label{na}\\
{\mbox {Type  II$_{\pm}$}}: && \dd
s^2=-\frac{\dd t^2}{G\mp \rrr^2\left(
1-D\rrr^2\pm\sqrt{(1-D\rrr^2)^2-K\rrr^4}
\right)^{-1}} \label{nb}\\
&&\quad  
+\frac{2\left( 1-D\rrr^2\pm\sqrt{(1-D\rrr^2)^2-K\rrr^4}
\right)}{\beta^2\left((1-D\rrr^2)^2-K\rrr^4
\right)}\,\dd\rrr^2 +\rrr^2(\dd\vartheta^2+\sin^2\vartheta\,\dd\varphi^2)
\nonumber
\eea
where $D$, $G$ and $K$ are real constants, $\beta$ is a positive rational  number,
$\rrr$ is a radial coordinate restricted to certain interval $]\rrr_1,\rrr_2[$,
$(\vartheta,\varphi)$  are spherical coordinates on $\mathbb S^2$ and $t\in \RR$.

Bertrand spacetimes are interesting in their own right because of their
definition in terms of closed orbits. From the point of view of
manifold theory, closed geodesics have long played a preponderant role in
Riemannian geometry~\cite{Kl83}. A somewhat similar question, that of
characterizing all Riemannian manifolds whose geodesics are all closed,
is still wide open~\cite{Be78,Be87}. Periodic trajectories are also
central in the theory of dynamical systems, where a generic dynamical
system with a compact invariant set does have many closed
orbits~\cite{PM82}. Obstructions do arise, of course, when
considering regular vector fields  whose orbits are all circles.

The aim of this paper is to prove that, given any $(3+1)$D Bertrand spacetime with Lorentzian metric $\eta$ (\ref{bs}), the term $V(r)$ can {\em always}  be  interpreted as either the intrinsic KC or the harmonic oscillator potential on the corresponding 3D Riemannian Bertrand space $(M,g)$. Therefore, we do not only have a result on closed trajectories of spherically symmetric spaces that extends Bertrand's theorem to 3D curved manifolds, but we also recover the key role played by the corresponding KC and harmonic oscillator potentials of these 3D spaces. 

In this respect, we stress that Perlick, in a purely relativistic setting,  proved that  KC-like and   oscillator-like $(3+1)$D metrics can only  be identified provided one considers, in this order, Bertrand spacetimes given by the values $\beta=1$ and $\beta=2$~\cite{perlick}. Our results show that a proper (non-relativistic) Kepler/oscillator potential can always  be established for any 3D Bertrand space $(M,g)$, independently of the value of $\beta$, and such Kepler/oscillator potentials give all the possible $V(r)$ terms in Perlick's classification of Bertrand spacetimes.

Furthermore, we discuss  explicitly some particular Bertrand  spaces of interest, such as the three classical Riemannian spaces of constant curvature~\cite{Doub} (this case was briefly pointed out by Perlick~\cite{perlick}), a Darboux space with non-constant curvature~\cite{PhysD}  as well as the spaces introduced by Iwai and Katayama~\cite{IK94,IK95} as a generalization of the MIC--Kepler~\cite{mica, micb, micc} and  Taub--NUT metrics~\cite{IK93}. We recall that the Euclidean Taub--NUT metric has
attracted considerable attention in the physics community (see, for instance,~\cite{Ma82,AH85, BCJ, BCJM, JL, GW07} and references therein) since, in particular, the relative motion of two monopoles in this metric is
asymptotically described by its geodesics~\cite{CFH90}. Moreover, the reduction of the Euclidean
Taub--NUT space by its $\rm{U}(1)$-symmetry, as performed in the analysis
of its geodesics, essentially gives rise to a specific Iwai--Katayama space~\cite{IK94}.  In this way, all of these different systems are unified within our approach in a common Bertrand space framework.

The article is organized as follows. In section 2 we present the metric structure of the 3D  Bertrand spaces by considering them within a particular class of spherically symmetric Riemannian spaces. In section 3 we identify some physically relevant examples of Bertrand  spaces. The construction (and interpretation) of the intrinsic KC and harmonic oscillator   potentials on generic  Bertrand spaces is addressed in section 4.   In section 5,  we explicitly study   the KC and oscillator potentials for the metrics described in section 3.
Some comments concerning the generalization of the results here presented and  their connection with  superintegrability properties of the associated Hamiltonian flows are drawn in the last section.


\sect{Bertrand spaces} \label{S:Bertrand}

Each of the (Lorentzian) Perlick metrics~\cite{perlick}   defines the kinetic energy on the
$(3+1)$D manifold $\mathcal M$. Alternatively, this construction provides a
natural Hamiltonian on the 3D spherically symmetric space $(M,g)$ with coordinates $(r,\vartheta,\varphi)$ and with
a central potential ${V}(r)$ 
given by   the denominator of the term in $\dd
t^2$ (\ref{aaa}). Therefore, according to (\ref{na}) and (\ref{nb}), hereafter we shall consider the Bertrand spaces defined  by
\bea
{\mbox {Type  I} }: && \dd
s^2=\frac{ \dd\rrr^2 }{\beta^2\left( 1+K\rrr^2\right) }
+\rrr^2(\dd\vartheta^2+\sin^2\vartheta\,\dd\varphi^2)
\label{nas}\\
{\mbox {Type  II$_{\pm}$}}: && \dd
s^2=\frac{2\left( 1-D\rrr^2\pm\sqrt{(1-D\rrr^2)^2-K\rrr^4}
\right)}{\beta^2\left((1-D\rrr^2)^2-K\rrr^4
\right)}\,\dd\rrr^2 +\rrr^2(\dd\vartheta^2+\sin^2\vartheta\,\dd\varphi^2) .
\nonumber\\
\label{nbs}
\eea

Consequently,  Bertrand spaces  are included within the large family of  3D spherically symmetric spaces. Their corresponding general metric is usually written in terms of a `conformal' factor depending on a radial coordinate, say $\ra$  
(different  from $r$), 
and sometimes in terms of its logarithm ($\rho=\ln\ra$). Hence in order to be able to relate our results with others given in the literature
 we shall consider  the following generic metric in a 3D Riemannian manifold $M$ written in  different coordinate systems:
\bea
&&\dd s^2=f(|\bq|)^2\,\dd\bq^2=f(\ra)^2(\dd \ra^2+\ra^2\dd\Om^2)\nonumber \\
&&\quad\ \ =F(\rho)^2\,(\dd
\rho^2+\dd\Om^2)=g(\rrr)^2 
\dd\rrr^2+\rrr^2\dd\Om^2
\label{metric}
\eea
where $|\bq|^2=\bq^2=\sum_{i=1}^3 q_i^2$ and $\dd\bq^2 =\sum_{i=1}^3\dd q_i^2$;
$f(|\bq|)\equiv  f(\ra)$, $F(\rho)$ and $g(\rrr)$ are   smooth functions; and  $\dd\Om^2=\dd\vartheta^2+\sin^2\vartheta\,\dd\varphi^2$ is the  standard metric on the unit $2$D sphere $\mathbb S^{2}$. The relations between the coordinate systems, $(q_1,q_2,q_3)$, $(\ra, \vartheta, \varphi )$, $(\rho, \vartheta, \varphi )$ and $(\rrr, \vartheta, \varphi )$ are given by 
  \bea 
 q_1&\!\!=\!\!&\ra \cos \vartheta      \qquad
q_2=\ra     \sin \vartheta \cos \varphi  \qquad 
  q_3 =\ra \sin \vartheta  \sin \varphi  \nonumber\\
    |\bq|&\!\!=\!\!&\ra \qquad \rho=\ln \ra \qquad F(\rho)=\ra f(\ra)\qquad \rrr=F(\rho)\qquad 
\dd\rho=\rrr^{-1} g(\rrr)\dd\rrr  .
\label{otrasmetrics}
\eea

 The scalar curvature $R$ of the   metric 
(\ref{metric}) is always determined  by the `radial' (but not  necessarily geodesic!) coordinate $\ra \leftrightarrow \rho  \leftrightarrow  \rrr $. This  curvature   is   generically non-constant and turns out to be
\bea
&&R=2\left( \frac{  f'(\ra)^2- 2 f(\ra)  \left(   
f''(\ra)+2 \ra^{-1}f'(\ra)  \right)}   {f(\ra)^4  } \right)  \cr
&& =2\left( \frac{   F'(\rho)^2+ F(\rho)^2 -2  F(\rho) F''(\rho) 
}   {F(\rho)^4    } \right) =2\left( \frac{ g(\rrr)^2+2\rrr g'(\rrr)/g(\rrr)-1}{\rrr^2g(\rrr)^2}
\right)
\label{curv}
\eea
where a prime denotes  derivative  with respect  
to  the corresponding  argument,  i.e.\  $f'=\dd f/\dd \ra $  and so on.


\subsect{Spaces of type I}

 If we now apply  the above results to the Bertrand  metrics of type I  
   (\ref{nas})  we find that
\be
g(\rrr)=\frac{1}{\beta\sqrt{ 
1+K\rrr^2}}
\qquad
\dd\rho= \frac{ \dd\rrr }{\beta\rrr\sqrt{
1+K\rrr^2} } 
\label{nddd}
\ee
which yield
\be
\rho=\frac 1\beta\,\ln\left(\frac{\rrr}{1+\sqrt{
1+K\rrr^2} } \right)
\quad
\mbox{or}\quad \ra= \left(\frac{\rrr}{1+\sqrt{
1+K\rrr^2} } \right)^{1/\beta}.
\label{ne}
\ee
Thus the Perlick  radial coordinate $r$ is given in terms of the `conformal' ones by
\be
\rrr =\frac{2}{{\ra}^{-\beta}-K {\ra}^{\beta} } = \frac{2}{{\rm e}^{-\beta\rho}-K {\rm e}^{\beta\rho} } .
\label{nee}
\ee
Hence  the   metric (\ref{nas})   can also be written as
\be
\dd s^2=\frac{4}{(\ra^{-\beta}-K \ra^{\beta})^2 }\left(\dd \ra^2+\ra^2\dd\Om^2  \right) =\frac{4}{({\rm e}^{-\beta\rho}-K {\rm e}^{\beta\rho})^2 }\left(\dd\rho^2+
\dd\Om^2  \right) 
\label{nf}
\ee
and its scalar curvature  reads
\bea
&&R= -\frac 12\left\{ (\beta^2-1)(K^2 {\ra}^{2\beta}+{\ra}^{-2\beta} )+2 K (1+5\beta^2) \right\} \nonumber\\
&&\quad =-\frac 12\left\{ (\beta^2-1)(K^2 {\rm e}^{2\beta\rho}+{\rm
e}^{-2\beta\rho} )+2 K (1+5\beta^2) \right\} 
 \nonumber\\
&&\quad =-  \frac{2}{\rrr^2}\left\{  3\beta^2 K\rrr^2   +\beta^2-1\right\} .
\label{ng}
\eea


\subsect{Spaces of type II$_{\pm}$}

As far as the Bertrand  type II$_\pm$ is concerned,    the function $g(\rrr)$  is given by
\be
g(\rrr)=\frac{\sqrt{2}}{\beta}\left(\frac{ 1-D\rrr^2\pm\sqrt{(1-D\rrr^2)^2-K\rrr^4}  
 }{ (1-D\rrr^2)^2-K\rrr^4  }\right)^{1/2} .
\label{oa}
\ee
 The explicit form of $\rho(\rrr)$   can be obtained by  taking into account that:  
 $$
\rho(r)=\int^r  \frac{g(\rrr) }{r}\dd\rrr=\int^r r\,   \frac{g(\rrr) }{r^2} \dd\rrr 
=r\int^r  \frac{g(\rrr) }{r^2} \dd\rrr  -\int^r  \left( \int^r  \frac{g(r) }{r^2} \dd\rrr\right)\dd r 
$$
which yields 
\be
\rho(r)=r U(r)-\int^r U(r)\dd r
\ee
where
\be
U(\rrr)=\int^r  \frac{g(\rrr) }{r^2} \dd\rrr =\mp \frac{\sqrt{2}} {\beta\rrr} \left( 1-D\rrr^2\pm\sqrt{(1-D\rrr^2)^2-K\rrr^4}
\right)^{1/2} .
\label{oc}
\ee
The integral $\int^r U(r)\dd r$, although it can be explicitly written, is quite involved, so that we omit it and   present the results for this family of metrics just in terms of $\rrr$.

The scalar curvature   turns out to be
\be
R=\frac{3}{\rrr^2}\left\{\frac 23 (1-\beta^2) +\beta^2  
\frac{ (K -D^2)\rrr^4+1 }{1-D\rrr^2\pm\sqrt{(1-D\rrr^2)^2-K\rrr^4} }
\right\}.
\label{ob}
\ee


\sect{Examples of Bertrand spaces}

In this section we identify within the common framework of the Bertrand spaces (\ref{nas}) and (\ref{nbs})
several relevant specific cases that have been introduced in the literature by using rather different approaches. In particular, we show that the three classical Riemannian spaces of constant curvature~\cite{Doub}, a generalization of the Darboux surface of type III~\cite{PhysD} and the spaces of non-constant curvature studied by Iwai--Katayama~\cite{IK94,IK95} belong to the family of Bertrand spaces.


\subsect{Spaces with constant curvature}

So far we have explicitly shown that Bertrand spaces are generically of non-constant curvature. Nevertheless the three classical Riemannian spaces with constant
sectional curvature $\k$, the spherical   $(\k>0)$, Euclidean
  $(\k=0)$, and hyperbolic  spaces $(\k<0)$, can be recovered as special instances of Bertrand spaces. The metric of such  spaces of constant curvature can  be collectively written
  in terms of  Poincar\'e coordinates $\bq\in\RR^3$ (coming from   a  stereographic
projection in $\RR^{4}$~\cite{Doub}) or in geodesic polar coordinates $(\rr,\vartheta, \varphi )$ as~\cite{VulpiLett,BHletter}
   \be
   \dd s^2=\frac{4}{(1+\k
\>q^2)^2}\,\dd\>q^2=
   \dd\rr^2 + \frac{\sin^2( \sqrt{\k}\, \rr) } {\k}\, \dd\Om^2.
  \label{wa}
  \ee
We remark   that $\rr$ is the distance along a minimal geodesic that joins the particle and the origin in our  Riemannian  3-space; this geodesic  distance does not coincide   neither with  $\ra=|\bq|$ nor with $r$ (\ref{otrasmetrics}).  By taking into account that the conformal function of the metric (\ref{wa}) is 
$f(\ra)=2/  (1+\k
\ra ^2)$, and  by defining
\be
\ra  = \frac{1}{\sqrt{\k}}\,\tan\left(\sqrt{\k}\,\frac \rr 2\right)\quad \mbox{or}\quad
r  = \frac{1}{\sqrt{\k}}\sin\left(\sqrt{\k}\ \rr  \right)
\label{wb}
\ee
we find  that the generic metric (\ref{metric}) reduces to (\ref{wa}).

This result directly comes from   Bertrand spaces (\ref{nas})  and   (\ref{nbs}) by setting~\cite{perlick}:
\be
{\mbox {Type  I} }: \    \beta=1   \quad K=-\k\qquad
{\mbox {Type  II}}_+ : \   \beta=2  \quad K=0 \quad D=\k 
\label{wwb}
\ee
which yields
\be
 \dd s^2=\frac{ \dd\rrr^2 }{ 1-\k\rrr^2}
+\rrr^2\dd \Om^2
\label{wc}
\ee
that is, $g(r)=(1-\k r^2)^{-1/2}$.
By applying the change of radial coordinates (\ref{wb}), we obtain that the latter metric coincides with (\ref{wa}).
Note also that the scalar curvature (\ref{ng})  is constant and equal to $6\kappa$.


\subsect{Darboux space of type III}

 The so-called  2D \emph{Darboux spaces} were studied by Koenigs~\cite{Ko72} and these are the only surfaces with non-constant curvature admitting two functionally independent integrals, quadratic in the momenta, that commute with the geodesic flow~ \cite{Ko72,KKMW02,KKMW03,pogoa,pogob}. There are  four types of such spaces. We shall focus on the Darboux surface of type III, $\cD_{\rm III}$, whose metric in terms of isothermal coordinates~\cite{Do76}
$(u,v)$ is given by
\be
\dd s^2={\e^{-2u}}(1+\e^u)  \,(\dd u^2+\dd v^2) .
\label{wd}
\ee
 If we define new coordinates $(q_1,q_2)$ through
 \be
 q_1=\e^{-u/2}\cos (v/2)\qquad 
 q_2=\e^{-u/2}\sin (v/2)
 \ee
 the metric (\ref{wd}) is transformed, up to a constant factor, into
 \be
 \dd s^2= (1+q_1^2+q_2^2)(\dd q_1^2+\dd q_2^2).
 \label{we}
 \ee
 From this expression an $N$D generalization of $\cD_{\rm III}$ was recently proposed in~\cite{PhysD}. Here we consider the 3D version given by
\be
 \dd s^2= (\kk+\bq^2)\dd \bq^2
 \label{wf}
 \ee
where $\kk$ is an arbitrary real constant; this metric is clearly of the form (\ref{metric}) with
\be
f(\ra)=\sqrt{\kk +\ra^2}\qquad r=F(\rho)=\e^\rho\sqrt{\kk+\e^{2\rho}}.
\label{wg}
 \ee
Hence
\be
  \rho=\frac 12 \ln\left( \frac{-\kk \pm\sqrt{\kk^2+4 r^2}}{2}\right)
\label{wh}
 \ee
which gives
\be
g(r)=r\,\frac{\dd\rho}{\dd r}= \frac{ \sqrt{\kk^2+4 r^2} \pm \kk   }{2\sqrt{\kk^2+4 r^2}} 
\label{wi}
 \ee
 so that the metric (\ref{wf}) is transformed into
 \be
  \dd s^2= \frac{\kk^2+2 r^2\pm \kk \sqrt{\kk^2+4 r^2}}{2(\kk^2+4 r^2)}\,\dd r^2+r^2\dd\Om^2 .
  \label{wff}
 \ee
 In this way  we show  that 
$\cD_{\rm III}$ is in fact a Bertrand space of type II$_{\pm}$ (\ref{nbs})  provided that the Perlick  parameters
are taken as 
\be
\beta=2\qquad K=D^2 \qquad D=-2/\kk^2 .
\label{wjj}
\ee
The corresponding scalar curvature reads
\be
R=-6\,\frac{2\kk+\ra^2}{(\kk+\ra^2)^3}=-6\,\frac{2\kk+\e^{2\rho}}{(\kk+\e^{2\rho})^3}
=3\,\frac{\kk^4+2\kk^2 r^2-2 r^4\mp\kk^3\sqrt{\kk^2+4 r^2} }{r^6} .
\label{wj}
\ee


\subsect{Iwai--Katayama spaces}

These are the spaces underlying the so-called  `multifold Kepler' systems introduced by these authors in~\cite{IK95}. Their metric is given by
\be
\dd s^2=\ra^{\frac  1\nu -2} (a + b\,\ra^{1/\nu})  \left( \dd \ra^2+\ra^2\dd\Om^2\right)  
 \label{wk}
\ee
where $a$ and $b$ are two real constants, while $\nu$ is a rational number. This metric is of the form (\ref{metric}) provided that
\be
f(\ra)=\ra^{\frac  1{2\nu} -1} (a + b\,\ra^{1/\nu})^{1/2}\qquad 
r=\ F(\rho)=\left(a \exp(\rho/\nu) + b \exp (2\rho/\nu)\right)^{1/2}
\label{wl}
\ee
so that
\be
\rho= \nu \ln\left( \frac{-a \pm\sqrt{a^2+4 b r^2}}{2 b}\right) 
\label{wm}
\ee
with $b\ne 0$. From it we compute  the Perlick function $g(r)$:
\be
g(r)=r\,\frac{\dd\rho}{\dd r}= \nu \, \frac{ \sqrt{a^2+4 b r^2} \pm a   }{\sqrt{a^2+4 b r^2}} .
\label{wn}
 \ee
Therefore the Iwai--Katayama  metric (\ref{wk}) can be written as
\be
\dd s^2=2\nu^2\,\frac{a^2+ 2 b r^2 \pm a \sqrt{a^2+4 b r^2} }{a^2+4 b r^2}
\,\dd r^2+r^2\dd\Om^2 
 \label{wo}
\ee
which is proven to belong to the Bertrand family II$_\pm$ (\ref{nbs}) under the identifications
 \be 
 \beta^2 = {\frac{1}{\nu^2}} \qquad K = D^2  \qquad D= -{\frac{2b}{a^2}} .
 \label{relations}
 \ee
  Hence, the `multifold Kepler'  metric  is actually a subcase of the Perlick metric.
  The scalar curvature for these spaces turns out to be
 \bea
 &&R=\frac{ 8 a b (\nu^2-1)+ 4 b^2(\nu^2-1)  \ra^{1/\nu}  +a^2 (4\nu^2-1) \ra^{-1/\nu}}{2\nu^2(a+b \ra^{1/\nu})^3}\nonumber\\
 &&\quad =\frac{ 3 a^4+6 a^2 b r^2+ 8 b^2 (\nu^2-1) r^4\mp 3 a^3\sqrt{a^2+4 b r^2} }{4 \nu^2 b^2 r^6} 
\eea  
which acquires a simpler expression once the parameter $D$ (\ref{relations})  is plugged in:
\be 
R = \frac{2( \nu^2-1)}{\nu^2 r^2} + 3\, \frac{ 1-Dr^2 \mp \sqrt{ 1-2Dr^2} }{ \nu^2 D^2 r^6}.
\ee

We stress that the Darboux space of type III with metric (\ref{wff}) is a particular case of these 
 Iwai--Katayama spaces as it can be recovered by setting $b=1$, $a=\kk$ and $\nu=1/2$. 
  Nevertheless a separate study of $\cD_{\rm III}$ allows us to elaborate on the connections with Darboux surfaces and gives rise to a neater analysis of the relationships  with KC and harmonic oscillator potentials.


\sect{KC  and oscillator  potentials on Bertrand spaces}

It is well known that the KC potential $1/r$ ($r^2=\>q^2=q_1^2+q_2^2+q_3^2$) in $\RR^3$ is simply a multiple of the (minimal) Green function of the Laplacian $\De_{\RR^3}:=\sum_i\dif{^2}{q_i^2}$, whereas the harmonic oscillator potential is the inverse of its square. We shall   extend this prescription, 
 holding on the flat Euclidean space, to  
 a  generic curved manifold $M$ with metric (\ref{metric}).  As we shall see, such a prescription is fully consistent with all results previously known in the literature.

The Laplace--Beltrami operator on $(M,g)$ can be written as
\be
\De_{M}=\frac1{\sqrt g}\sum_{i,j=1}^3\dif{}{q_i}\sqrt g\,g^{ij}\dif{}{q_j}
\ee
in terms of an arbitrary set of local coordinates $\bq=(q_1,q_2,q_3)$. In this formula, as usual, $g^{ij}$ denotes the inverse of the matrix tensor $g_{ij}$ in these coordinates and $g$ is the determinant of $g_{ij}$. Let
us define the radial function $U(|\>q|)$, which can be expressed in terms on any of the radial coordinates $\ra$, $\rho$ or $r$ given in (\ref{otrasmetrics}), as the positive non-constant solution to the equation
\be
\De_{M}U=0\quad\mbox{on}\quad M\minus\{\textbf0\} .
\ee
A short calculation
using~(\ref{metric}) and (\ref{otrasmetrics}) shows that
\bea
&&\Delta_{M}U =
\frac{ 1}{\ra^2 f(\ra)^3}\frac\dd{\dd
\ra}\left(\ra^2 f(\ra)\frac{\dd U(\ra)}{\dd
\ra}\right) =
\frac{ 1}{F(\rho)^3}\frac\dd{\dd
\rho}\left(F(\rho)\frac{\dd U(\rho)}{\dd
\rho}\right) \nonumber\\
&&\qquad    \  \ =\frac{ 1}{r^2 g(r)}\frac\dd{\dd
r}\left(  \frac{r^2}{g(r)}\frac{\dd U(r)}{\dd
r}\right) .
\eea
This means that the symmetric Green function $U$ of the Laplace--Beltrami
operator in $M$~(see~\cite{LT87,EP06d} and
references therein)    is given by
\begin{equation}
U =\int^\ra\frac{\dd \ra}{\ra^2f(\ra)}=\int^\rho\frac{\dd \rho}{F(\rho)}
=\int^\rrr\frac{g(\rrr)}{\rrr^2}\,\dd\rrr
\label{nnd}
\end{equation}
up to inessential additive and multiplicative constants.
Its
potential-theoretic interpretation now leads to define the
\emph{intrinsic KC potential} on the $3$D manifold $M$ as
\begin{equation}\label{kc}
{\cal U}\Kep :=\cte\,U 
\end{equation}
where  $\cte$ is an arbitrary constant. The
\emph{intrinsic harmonic oscillator potential} in $M$ is defined to be proportional
to the inverse square of the KC potential:
\begin{equation}\label{harm}
{\cal U}\Harm :=\frac \cte{U^2}\,.
\end{equation}

\noindent
In what follows we  apply these results to the Perlick  metrics (\ref{nas}) and
(\ref{nbs}).


\subsect{Type I: intrinsic KC potential}

By considering (\ref{nf}) and  (\ref{nnd})
 we obtain the
Green function  $U$ corresponding to the Bertrand  space of type I with metric  (\ref{nas}):
\be
U= -\frac 1{2\beta}\left( {\ra}^{-\beta}+K {\ra}^{\beta}
\right) = -\frac 1{2\beta}\left( {\rm e}^{-\beta\rho}+K {\rm e}^{\beta\rho}
\right)  = - \frac 1\beta\,\sqrt{\rrr^{-2}+K} .
\label{nh}
 \ee
Then the intrinsic KC and harmonic oscillator potentials on these spaces are defined by
\bea
&&{\cal U}\Kep = -\frac \alpha{2\beta}\left( {\ra}^{-\beta}+K {\ra}^{\beta}
\right) = -\frac \alpha{2\beta}\left( {\rm e}^{-\beta\rho}+K {\rm e}^{\beta\rho}
\right)  = - \frac \alpha\beta\,\sqrt{\rrr^{-2}+K} 
\label{nha}\\
&&{\cal U}\Harm =\cte \,\frac {4\beta^2} {\left( {\ra}^{-\beta}+K {\ra}^{\beta}
\right)^2}=\cte \,\frac {4\beta^2} {\left( {\rm e}^{-\beta\rho}+K {\rm
e}^{\beta\rho}
\right)^2}
=\cte \,\frac {\beta^2} {\rrr^{-2}+K} .
\label{nhb} 
\eea
Next if we compare these expressions with the warping function appearing in the $(3+1)$D Bertrand spacetime (\ref{na}),
\be
V(r)=G+\sqrt{\rrr^{-2}+K}  
\label{pot1}
\ee
 we find that this is exactly the  {\em intrinsic KC potential} on the Bertrand space (\ref{nas}), $V= {\cal U}\Kep +G$, with $\alpha=-\beta$ and  $G$ playing the role of an additive constant of the potential.

\subsect{Type II$_{\pm}$: intrinsic harmonic oscillator potential}

In this case, the Green function $U(r)$ has exactly the expression (\ref{oc}). Hence 
the intrinsic KC and harmonic oscillator potentials on the curved spaces   of type
II$_{\pm}$ are given (only in terms of $r$) by
 \bea
&&{\cal U}\Kep(r)=\mp\cte\frac{\sqrt{2}} {\beta\rrr} \left(
1-D\rrr^2\pm\sqrt{(1-D\rrr^2)^2-K\rrr^4}
\right)^{1/2}  \label{od}\\
&&{\cal U}\Harm(r)=
\frac{\cte\beta^2\rrr^2}{2\left(1-D\rrr^2\pm\sqrt{(1-D\rrr^2)^2-K\rrr^4}\right)} .  
\label{odd}
\eea

 By taking into account the equations (\ref{nb}) and (\ref{odd}) we conclude that the
underlying potential of the $(3+1)$D Bertrand spacetimes of type II$_{\pm}$, namely
 \be
V(r)= {G\mp \rrr^2\left(
1-D\rrr^2\pm\sqrt{(1-D\rrr^2)^2-K\rrr^4}
\right)^{-1}}
\label{pot2}
\ee
 is exactly the {\em intrinsic harmonic oscillator}  on the corresponding $3$D
Bertrand space (\ref{nbs}), $V= {\cal U}\Harm+G$, provided that $\alpha=\mp 2/\beta^2$ and  $G$ being again  an additive constant of the potential.

 At this point some comments concerning the above results and those obtained by Perlick~\cite{perlick} seem to be pertinent. So far we have obtained  all the functions $V(r)$ in the families  I and II$_\pm$ of $(3+1)$D Bertrand spacetimes as, respectively, the intrinsic KC and oscillator potentials on the corresponding 3D Bertrand spaces. This identification holds for any value of $\beta$, 
while in the exhaustive analysis performed by Perlick it is shown that relativistic analogues of the KC and oscillator systems can only be obtained by considering physical constraints that lead to fixed values  $\beta=1$ and $\beta=2$, respectively, of the Bertrand spacetimes (cf.~propositions 4 and 5 in~\cite{perlick}). Note that, obviously, both results are not in contradiction since the definition of the Kepler/oscillator potentials on 3D Bertrand spaces is based on a different (non-relativistic) physical setting that does not exclude any value of $\beta$.


\sect{Examples of Bertrand spacetimes}

In this section we firstly specialize the intrinsic KC and harmonic oscillator potentials for the particular Bertrand spaces described in section 3 and, secondly, we single out those Bertrand spacetimes associated to them.


\subsect{Bertrand spacetimes from constant curvature  spaces}

Let us consider the   classical Riemannian spaces of constant sectional curvature $\k$ described in section 3.1 with metric  (\ref{wa}) or   (\ref{wc}). The Green function (\ref{nnd}) is therefore given by
\be
U=-\frac{1-\k \ra^2}{2\ra}=-\frac{\sqrt{1-\k r^2}}{r} .
\ee
The definition of the geodesic radial coordinate $\rr$   (\ref{wb})   gives
 \be
 \frac{\tan( \sqrt{\k}\, \rr) }{\sqrt{\k}}= \frac{r}{\sqrt{1-\k r^2}} .
 \ee
In this way the corresponding KC and oscillator potentials  are found to be
 \bea
{\mbox {Type  I} }: &&    {\cal U}\Kep =  -\alpha\,\frac{1-\k \ra^2}{2\ra}=-\alpha \sqrt{\rrr^{-2}-\k}  
= -\alpha\,\frac{\sqrt{\k}}{\tan( \sqrt{\k}\, \rr) }
\label{ggf}\\
{\mbox {Type  II}}_+ : &&
 {\cal U}\Harm  =  \frac{4 \alpha\ra^2}{(1-\k \ra^2)^2}=           \frac{\alpha \rrr^2} {1-\k\rrr^2}  
 = \alpha\,\frac{\tan^2( \sqrt{\k}\, \rr) }{ {\k}}
\label{ggh}
\eea
which are, in this order,    particular cases of the Bertrand potential of type I (\ref{nha})  and of the    type  II$_{+}$ (\ref{odd}) provided that the relations (\ref{wwb}) have been introduced. Obviously we can always add the additive constant $G$.   Notice that,  within this framework, we have recovered the   well known expressions for these potentials (see~\cite{VulpiLett, BHletter} and references therein).

Consequently both the KC and harmonic oscillator potentials on these spaces of constant curvature lead to
particular cases of Bertrand spacetimes; these are
 \bea
{\mbox {Type  I} }: &&  \dd s^2=-\frac{\dd t^2}{G+ \sqrt{ \rrr^{-2} -\k} } +\frac{ \dd\rrr^2 }{ 1-\k\rrr^2}
+\rrr^2\dd \Om^2
\label{ggff}\\
{\mbox {Type  II}}_+ : &&\dd s^2=-\frac{\dd t^2}{G-r^2\left( 2(1-\k \rrr^{2} ) \right)^{-1}  }  +\frac{ \dd\rrr^2 }{ 1-\k\rrr^2}
+\rrr^2\dd \Om^2\label{gghh}
\eea
where we have  
fixed 
 $\alpha=-1$ and $\alpha=-1/2$, respectively.


\subsect{A Bertrand--Darboux spacetime}

In this case the Green function (\ref{nnd}) of the metric (\ref{wf}) or (\ref{wff}) turns out to be
\be
U=-\frac{\sqrt{\kk+\ra^2}}{\kk\ra}=-\frac{\sqrt{1+\kk\, \e^{-2\rho}} }{\kk}
=-\frac{\kk\pm\sqrt{\kk^2+4 r^2} }{2\kk r}.
 \ee
By using this Green function we can define    the intrinsic KC and oscillator potentials on $\cD_{\rm III}$. However since such a space   appears as a particular Bertrand space of type II$_\pm$, {\em only} the oscillator potential 
can be considered in this context; namely
\be
{\mbox {Type  II}}_\pm: \quad
 {\cal U}\Harm  = \frac{\alpha  \kk^2\ra^2}  {{\kk+\ra^2}}=
\frac{\alpha  \kk^2}{ {1+\kk\, \e^{-2\rho}} } = \frac{2\alpha \kk^2 r^2}{\kk^2+2 r^2\pm \kk \sqrt{\kk^2+4 r^2}}
\ee
which is reproduced from (\ref{odd}) through the identifications (\ref{wjj}). We recall that this was exactly the oscillator potential on $\cD_{\rm III}$ worked out in~\cite{PhysD} in terms of the radial coordinate $\ra=|\>q|$. Thus its corresponding Bertrand spacetime of type   II$_{\pm}$ reads
\be
 \dd s^2= -\frac{\dd t^2}{G\mp \kk^2 r^2\left(\kk^2+2 r^2\pm \kk \sqrt{\kk^2+4 r^2} \right)^{-1}}+\frac{\kk^2+2 r^2\pm \kk \sqrt{\kk^2+4 r^2}}{2(\kk^2+4 r^2)}\,\dd r^2+r^2\dd\Om^2 
 \ee
provided that $\alpha=\mp 1/2$.


\subsect{Bertrand spacetimes from Iwai--Katayama spaces}

The  Iwai--Katayama metric (\ref{wk}) or (\ref{wo}) gives rise to the following Green function
\bea
&&U=-\frac{2\nu}{a}\sqrt{a \ra^{-1/\nu}+b}=-\frac{2\nu}{a}\sqrt{a \e^{-\rho/\nu}+b}\nonumber\\
&&\qquad =-\frac{\sqrt{2}\nu}{a r}\left(a^2+2 b r^2\pm a\sqrt{a^2+ 4 b r^2}      \right)^{1/2}
\label{xxzz}
\eea
which allows for the definition of the corresponding intrinsic KC and oscillator potentials. The results of section 3.3 show that only the oscillator potential can be constructed on these spaces in order to obtain a Bertrand spacetime of type II$_\pm$ by means of the relations  (\ref{relations}); explicitly
\bea
{\mbox {Type  II}}_\pm: &&  
 {\cal U}\Harm  =\frac{\alpha a^2}{4 \nu^2(a \ra^{-1/\nu}+b)}
 =\frac{\alpha a^2}{4 \nu^2(a  \e^{-\rho/\nu}+b)}  \nonumber \\
&&\quad  = \frac{\alpha a^2 r^2}{2\nu^2}\left(a^2+2 b r^2\pm a \sqrt{a^2+ 4 b r^2} \right)^{-1}.
\label{ssaa}
\eea

These results can be analysed in relation with the   so called `multifold Kepler' potential $V_{\rm IK}$ on  (\ref{wk})  introduced in~\cite{IK95} which  is given by
\be
V_{\rm IK}(\ra)=\frac { \ra^{2- \frac {1}{\nu}} }{ a + b \ra^{\frac 1\nu}}\left(
\mu^2 \ra^{-2}+ \mu^2 c\, \ra^{\frac 1\nu -2}+ \mu^2 d\,\ra^{\frac 2\nu -2} \right)
\ee
where $\mu$, $c$ and $d$ are real constants. We remark that this potential is of physical relevance as   for $\nu=1$ (together with   some specific values of the parameters $a$, $b$, $c$ and $d$ as pointed out in~\cite{IK95}) it   comprises the MIC--Kepler problem~\cite{mica,micb,micc,KN07, NY08} as well as the 
Taub-NUT one~\cite{Ma82,AH85, GW07,GM86,FH87,GR88}. 

Now notice that the  angular momentum of the system ${\bf L}^2$  conjugate to $\dd\Om^2$, appearing in the metric (\ref{wk}), is a constant of the motion (see e.g.~\cite{PhysD}) so that the term $\mu^2 \ra^{-2}$  can be reabsorbed within the kinetic energy term since ${\bf L}^2+\mu^2$ is obviously an integral of motion,  $\mu^2$ being an additive constant. Then the potential   reduces, up to the  multiplicative
constant $\mu^2$, to
\be
V_{\rm IK}(\ra)=\frac {   c +  d\,\ra^{\frac 1\nu }  
 }{ a + b \ra^{\frac 1\nu} } .
 \label{ssbb}
\ee
This, in turn, means that $V_{\rm IK}$ can be regarded as a M\"obius   map on the positive real line with variable $x=\ra^{1/\nu } =\exp (\rho/\nu)\in \RR^+$ under the action of the group $SL_2(\RR)$:
$$
x\to V_{\rm IK}=g\cdot x=\frac{d x + c}{b x + a}\quad \mbox{where}\quad
g= \left(\begin{array}{cc}
d&c\\
b&a
\end{array}\right)\in  SL_2(\RR) .
$$
Next if we add a constant $G$ to the oscillator potential (\ref{ssaa}) we find that
\be
 {\cal U}\Harm(\ra)  +G=\frac {   c +  d\,\ra^{\frac 1\nu }  
 }{ a + b \ra^{\frac 1\nu} } 
  \ee
  provided that
  \be
  G=c/a\qquad ad-bc=\frac{\alpha a^3}{4\nu^2} .
  \ee
   Note that   the second relation is a constraint between the parameters of the potential and those of the metric, which arises from the requirement that 
   the M\"obius transformation be non-degenerate.

Consequently the `multifold Kepler' potential $V_{\rm IK}$ (\ref{ssbb}) is actually an intrinsic harmonic oscillator (with constant $d$ plus an 
 additive constant $c$) on the  Iwai--Katayama spaces studied in section 3.3.

Finally, we write the resulting metric of the  associated Bertrand spacetimes (which are of type II$_\pm$):
\bea
&& \dd s^2=-\frac{\dd t^2}{G\mp a^2 r^2 \left(a^2+ 2 b r^2 \pm a \sqrt{a^2+4 b r^2}  \right)^{-1} }\nonumber\\
&&\qquad\quad +2\nu^2\,\frac{a^2+ 2 b r^2 \pm a \sqrt{a^2+4 b r^2} }{a^2+4 b r^2}
\,\dd r^2+r^2\dd\Om^2 
\eea
where we have set $\alpha=\mp 2\nu^2$.


\sect{Concluding remarks}

In this paper we have explicitly shown that the families of $(3+1)$D Bertrand spacetimes I and II$_\pm$ can be neatly understood in terms of the  `intrinsic' KC and harmonic oscillator potential on the corresponding 3D Bertrand spaces. In this sense,
Bertrand spaces can be simply described as 
those manifolds where the classical Bertrand's theorem still holds. Moreover, the resulting central potentials cannot be but the KC or the harmonic oscillator potentials. Our general description of Bertrand spaces and potentials has been also applied to some relevant cases which so far appeared in the literature as stemming from 
 different approaches.   In particular we have described  in  a unified way  the intrinsic oscillator potential  on  the Darboux III and Iwai--Katayama spaces. As a byproduct of these examples we have proven that 
the so called `multifold Kepler' potential, and therefore the MIC--Kepler  and
Taub-NUT problems, comes out in a natural way within a certain class of $(3+1)$D Bertrand spacetimes of type  II$_\pm$, thus suggesting that  such potential should be actually regarded as an intrinsic harmonic  oscillator. 

It is worth recalling that the main algebraic feature of the KC and oscillator potentials on the $N$D Euclidean space is the fact that both Hamiltonian systems are {\em maximally superintegrable}: this means that both systems admit a maximum number $(2\,N - 2)$ of functionally independent and globally defined constants of the motion that Poisson-commute with the Hamiltonian. Moreover, it is well known that all bounded orbits of any maximally superintegrable Hamiltonian system are periodic. Thus, one could conjecture that the KC and oscillator systems on generic Bertrand spaces should be maximally superintegrable. In order to prove this statement, the explicit form of the $2\cdot 3 - 2=4$ independent  integrals of the motion that would commute with  these Hamiltonian flows should be found.
Note that this result is already known in the literature for the `$\nu$-multifold Kepler systems' with any rational value of $\nu=1/\beta$~\cite{IK95}.  We point out that, again, this problem is quite different to the one consisting in the study  of the maximal superintegrability of the geodesic motion on the $(3+1)$D Bertrand spacetimes which is, in fact, solved (see proposition 6 in~\cite{perlick}) and 
imposes that  $\beta=1,2$.

We will report on this question elsewhere~\cite{conjecture}. For completeness, let us mention that this fits naturally into the framework of Hamiltonian systems endowed with an $sl(2)$ Poisson coalgebra symmetry~\cite{BR98,PLB, sigmaOrlando}. It turns out that this coalgebra symmetry provides the {\em quasi-maximal superintegrability} of any  spherically symmetric Hamiltonian system defined on a Bertrand spacetime ({\em i.e.} three of the above integrals). Therefore, in order to achieve the maximal superintegrability of the KC and oscillator potentials only one additional integral has to be found by direct methods. The maximal superintegrability of motion in the particular case of Iwai--Katayama spaces has already been established~\cite{GM86,FH87,GR88}, and in fact its (complexified) symmetry
algebra is isomorphic to that of the KC or harmonic
oscillator potentials. Incidentally, it is worth stressing that this problem is related to similar questions studied by Koenigs in connection with the maximal
superintegrability of geodesic flows on surfaces \cite{PhysD,Ko72,KKMW02,KKMW03, pogoa, pogob}.

As a final remark, we would like to mention that despite having  focused our discussion on the
$(3+1)$D case,  the results here presented can be easily generalized to $(N+1)$ dimensions by rewriting the spherical symmetry in terms of an underlying $sl(2)$ Poisson coalgebra invariance.


\section*{Acknowledgements}

This work was partially supported by the Spanish MEC and by the Junta de
Castilla y Le\'on under grants no.\    MTM2007-67389 (with EU-FEDER support) and VA013C05 
(A.B.\ and F.J.H.), by the Spanish DGI and Complutense University under grants no.\
FIS2005-00752 and~GR69/06-910556 (A.E.), and by the INFN--CICyT (O.R.).


\end{document}